\documentclass[11pt]{article}
\usepackage{epsfig}

\newcommand{\ket}[1]{\vert {#1}\rangle}
\newcommand{\op}[1]{\hat{#1}}
\newcommand{\phaseop}{\hat{S}_{\epsilon}^{\pi}}

\begin{document}

\title{A Novel Approach to Quantum Heuristics for Structured Database
Search}
\author{Brian M. Murphy \\ 
\small{Research Institute for Advanced Computer Science} \\ 
\small{NASA Ames Research Center} \\ 
\small{Moffett Field, CA 94035-1000} \\ \\ 
\small{Department of Physics} \\
\small{Stanford University} \\ 
\small{Stanford, CA 94305}}
\maketitle
\begin{abstract}
An algorithm for structured database searching is presented and used to
solve the set partition problem.  $O(n)$ oracle calls are required in
order to obtain a solution, but the probability that this solution is
optimal decreases exponentially with problem size.  Each oracle call is
followed by a measurement, implying that it is necessary to maintain
quantum coherence for only one oracle call at a time.
\end{abstract}
\section{Introduction}
Quantum computers \cite{Deutsch85} are thought to be able to solve some
problems more efficiently than classical computers.  The most important
quantum algorithm is the Grover search \cite{Grover97,Grover98}
because of its applicability to
solving important computational problems, such as NP-complete problems.
In fact, it has already been shown that a nested Grover search can be used
to solve the graph-coloring problem, which is NP-complete \cite{Cerf98}.  

Here I report a new approach to structured database search, and I apply it
to the set partition problem, which is NP-complete.  Through simulation, I
find in this application that the number of required oracle calls is fewer
than a random classical search, but more than an unstructred Grover search
(see figures 3 and 4).  However, it is only necessary to maintain
quantum coherence for a single oracle call at a time, unlike Grover
searches which require that
quantum coherence is maintained throughout the entire running time.  Since
the short coherence times of quantum systems is the biggest obstacle to
quantum computing \cite{Yamaguchi00}, this is an interesting result.  

I will start by presenting the general principles outlining the new
approach to structured database search.  I will then show how they can be
specifically applied to the set partition problem.

\section{Quantum Algorithm for Structured Database Search}
\subsection{Definitions of Quantum Operators}
In order to search a mathematically-specified database, our quantum
computer will need two registers.  One is for the index of the database
item ($n$ qubits) and the other is for the data associated with that index
(size requirements discussed later).  This is a
mathematically-specified database because the data associated with the
index is the result of a mathematical function when given the index as input.

The first operator to define 
is one that creates the database to be searched without making any
measurement on the system.  The most natural way to accomplish this is in
a two-step process.  First,   
\begin{equation} \hat{W}_{s}\equiv\hat{W}\otimes\hat{I}\end{equation} or equivalently,
\begin{eqnarray} \hat{W}_{s} \vert i \rangle \vert j
\rangle\equiv(\hat{W} \vert i \rangle )\vert j \rangle
\end{eqnarray}
where $\hat{W}$ is an $n$-bit Walsh-Hadamard transform \cite{Yoshi6}. 
\\ Second, if we take $P$ to be the mathematical function relating the
index of the database item to its data, we want to define a unitary
operator that implements this function: 
\begin{equation} \label{defP}
\hat{P} \vert i \rangle \vert j \rangle\equiv \vert i \rangle \vert
(j+P(i))modN \rangle\end{equation} 
where $modN$ allows for overflow in the second register
($N\neq 2^n$).  In an optimization problem, $P$ is a cost
function. Now, if we define
\begin{eqnarray} \hat{A}\equiv\hat{P}\cdot\hat{W}_{s}\end{eqnarray}
we have that \begin{eqnarray} \hat{A}\vert i \rangle\vert j\rangle =
\frac{1}{2^{\frac{n}{2}}}\sum_{k=0}^{2^n-1} (-1)^{i\oplus k}\vert
k\rangle\vert (j+P(k))modN\rangle \end{eqnarray}
where $\oplus$ is binary bit-wise addition.
These definitions imply that \begin{eqnarray} \hat{A}^{-1}
=\hat{W}_{s}^{-1}\cdot\hat{P}^{-1}=\hat{W}_{s}\cdot\hat{P}^{-1}\end{eqnarray}
where \begin{eqnarray}\hat{P}^{-1} \vert i \rangle \vert j \rangle\equiv \vert i \rangle \vert
(j+N-P(i))modN \rangle\end{eqnarray}  
From this we see that\begin{eqnarray} \hat{A}^{-1}\vert i \rangle\vert j\rangle =
\frac{1}{2^{\frac{n}{2}}}(\sum_{k=0}^{2^n-1} (-1)^{i\oplus k}\vert
k\rangle)\vert (j+N-P(i))modN\rangle \end{eqnarray}
Most importantly, 
\begin{eqnarray} \hat{A}\vert 0 \rangle\vert 0\rangle =
\frac{1}{2^{\frac{n}{2}}}\sum_{s=0}^{2^n-1} \vert
s\rangle\vert P(s)\rangle \end{eqnarray}
The following phase flip operator will also be needed: 
\begin{eqnarray} \label{eq:defS}\hat{S}_{\epsilon}^{\pi}\vert i\rangle\vert j\rangle\equiv
\left\{ \begin{array}{ll} -\vert i\rangle\vert j\rangle & j\le\epsilon \\
\;\;\,\vert i\rangle\vert j\rangle & j>\epsilon \end{array} \right. \end{eqnarray}
States $i$ for which $P(i)$ is less than or equal to $\epsilon$ are called
\emph{good}, and states $i$ for which $P(i)$ is greater than epsilon are
called \emph{nogood}.  In general, our target state will be any $i_0$,
for which $P(i_0)$ is a global minimum. 
\\ Finally,
\begin{eqnarray}\hat{D}\equiv\hat{A}^{-1}\cdot\hat{S}_{\epsilon}^{\pi}\cdot\hat{A}\end{eqnarray}
You may notice that $\hat{D}$ is actually part of the general Grover
operator \cite{Brassard98}, but the use here will be very intuitively
different, as I will discuss later.
\subsection{Implementation of Quantum Operators}
In order to use these quantum operators in an algorithm, it is necessary
to show that they are efficiently implementable.  $\op{W}_{s}$ is trivial.
$\op{P}$ relies only on addition and a function evalution of $P$.  Since
these are both efficient classically, they can be implemented efficiently
on a quantum computer \cite{Deutsch85}.  $\phaseop$ can be implemented with
the use of a single work bit.  First, add $(N-\epsilon)$ to the second
register and store overflow in the work bit.  Second, invert the work bit
and perform a conditional phase flip.  Finally, uncompute to clear the
work bit.  
\subsection{The Algorithm} \label{sec:alg}
Here I will present the steps of an algorithm for structured database
search, then proceed to fill in the missing details. 

1) Clear to $\vert 0\rangle\vert 0\rangle$ \qquad\qquad $\vert\psi\rangle=\vert
0\rangle\vert 0\rangle$

2) Apply $\hat{D} \qquad\qquad\qquad\;\;\, \vert\psi\rangle=(\vert 0\rangle
- \frac{2}{2^{\frac{n}{2}}}\sum_{good} \vert s\rangle )\vert 0\rangle$

3) Measure $\qquad\qquad\qquad\;\;\, \vert\psi\rangle=\vert m\rangle\vert
0\rangle$

4) Use $m$ to half the size of the database. 

5) Repeat steps $1-4$ until a solution is found.

\subsection{Using Measurement to Reduce Problem Size}
Step 4 of the above algorithm is the crucial step.  The idea is basically
as follows: $\hat{D}$ consists of three operators:  $\hat{A}$ creates a database, $\phaseop$ flips
phases of good states, and $\hat{A}^{-1}$ uncomputes.  $\phaseop$ flips phases
based on the value in the 2nd register, but the subsequent phase interference affects
what is measured in the 1st register.  Using $\ket{m}$, we want to deduce
which first register values were entangled to second register values less
than $\epsilon$.

\subsubsection{Viewpoint as a Quantum Oracle}
I have defined $\hat{D}$ as a series of quantum operators on two
registers.  However, in the context of solving computational problems with
quantum algorithms, it is important to understand that $\hat{D}$ can be
equivalently viewed as operations on just one quantum register with the
help of a 'quantum oracle.'
\begin{eqnarray} \hat{D} & = &
\hat{A}^{-1}\cdot\hat{S}_{\epsilon}^{\pi}\cdot\hat{A} \nonumber \\ 
& = &
\hat{W}_{s}\cdot\hat{P}^{-1}\cdot\hat{S}_{\epsilon}^{\pi}\cdot\hat{P}\cdot\hat{W}_{s}\nonumber \\ 
& = & \hat{W}_{s}\cdot\textrm{(quantum oracle)}\cdot\hat{W}_{s} \end{eqnarray}
In the first viewpoint, 
a function evaluation acts in parallel to entangle all possible inputs to their outputs, and a
phase operator flips the phases of target states based on their output
value.  Then, in uncomputation, the inverse function evaluation returns all values in the
second register to $\vert 0\rangle$ so that phase interference can occur
between different states in the first register.

In the quantum oracle viewpoint, only one quantum register is used
 \emph{explicitly}. The phases of certain target states are flipped by a 
quantum oracle that uses machinery whose details we do not examine.   

\subsubsection{Deterministic Measurement}
\label{sec:DetMeas}
So far, the only new idea I have presented is to propose that the steps in
section 3 could be considered as an algorithm.  The rest of what 
I have covered is basically a summary of how I understand and use existing
ideas.  Now I will begin to explain step 4 of
the algorithm and justify the claim that $\hat{D}$ can be used for
structured database search.

In general, the state $\op{D}\ket{0}\ket{0}$ will be a superposition of
many eigenstates of the computational basis, and when we measure this
state, the value we obtain for $m$ is not deterministic.

However, it is very useful to ask the following question: what would be
the structure of a problem instance for which $\op{D}\ket{0}\ket{0}$ is an
eigenstate of the computational basis?  This is an easy question to
answer with the quantum oracle viewpoint:

\begin{eqnarray} \op{W}\cdot\textrm{(oracle)}\cdot\op{W}\ket{0} & = &
\ket{x}\nonumber  \\ \Rightarrow \textrm{(oracle)}\cdot\op{W}\ket{0} & = &
\op{W}\ket{x} \\ \textrm{using}\;\; \op{W}^2 = \op{I}\nonumber \end{eqnarray}

From this we see that \emph{if} the best half of the states in our database
exactly corresponds to the half of the database whose phase is flipped in the
Walsh-Hadamard transformation of some $\ket{x}$, \emph{then} after measurement we
will obtain $\ket{m}=\ket{x}$ with certainty.

\subsubsection{Choosing a Subset} \label{sec:choose}
Based on the function $P$, we could divide our database of $2^n$ states into a
best half and worst half.  The 1st register values of the best half will
not be random, or else this would be an \emph{un}structured database
search.  Nonetheless, their structure could easily be sufficiently
complicated that we could not adjust $\op{A}$ to create a superposition of
only those states \cite{Barenco95}.  More importantly, if our goal is to
reduce the size of our database by a factor of 2, then it is sufficient to
choose any half that still contains the state $i_0$.
Therefore, it seems reasonable to look for a way to \emph{approximate} the
best half of solutions.  

I propose the following: after measuring and obtaining $\ket{m}$, keep the
$2^{n-1}$ items whose phases are flipped in the expansion of
$\op{W}\ket{m}$.  This will be the half of the database that we use to
approximate the best half, motivated by the finding in section
\ref{sec:DetMeas} that if this approximate half was really the best half, then we would
have measured $\ket{m}$ with certainty.

Details for exactly how these states are chosen will be given later in the
context of the set partition problem.  In this case we will see that it is
always possible to efficiently create a superposition of the database
items we want, but in general this may or may not be true. 
\subsubsection{Measurement Probabilities}
The above section implicity assumes that we can choose an $\epsilon$ such
that $\phaseop$ flips exactly half of the states of the database.
However, this is not a strict requirement.  

When measuring the state $\hat{D}\vert 0\rangle\vert 0\rangle$, what is
the probability of measuring a given state $\vert x\rangle$ in the first
register? (second register is deterministically $\vert 0\rangle$)

First, in order to help quantify the action of the oracle, define:
\begin{eqnarray}\label{eq:defTheta} \Theta_{\epsilon}(k)\equiv \left\{ \begin{array}{ll} -1 &
P(k)\le\epsilon \\ \;\;\, 1 & P(k) > \epsilon \end{array} \right. \end{eqnarray}

Now we can calculate the probability of measuring $\vert x\rangle$,
$M(x)$:
\begin{eqnarray} M(x) & \equiv & \arrowvert\langle x\vert\hat{D}\vert
0\rangle\arrowvert ^2 \\ \langle x\vert\hat{D}\vert 0\rangle & = & \langle
x\vert\hat{W}\cdot\textrm{oracle}\cdot\hat{W}\vert 0\rangle\nonumber \\ 
& = & \Big( \frac{1}{2^{\frac{n}{2}}}\sum_k \langle k\vert(-1)^{k\oplus
x}\Big) \Big(\frac{1}{2^{\frac{n}{2}}}\sum_k\Theta_{\epsilon}(k)\vert k\rangle\Big) \nonumber\\
& = & \frac{1}{2^n}\sum_{k} (-1)^{k\oplus x}\Theta_{\epsilon} (k) 
\end{eqnarray} 

Now, let

F = number of states $\vert k\rangle$ such that $(-1)^{k\oplus x}= -1$ and
$\Theta_{\epsilon}(k) = -1$

and

N = number of states $\vert k\rangle$ such that $(-1)^{k\oplus x}= 1$ and
$\Theta_{\epsilon}(k) = -1$

With a little algebra, \begin{eqnarray}\label{eq:probFN} \langle x\vert\hat{D}\vert
0\rangle = \frac{F-N}{2^{n-1}} \end{eqnarray}
\begin{eqnarray} \Rightarrow M(x) = \Big(\frac{F-N}{2^{n-1}}\Big)^2 
\end{eqnarray}
\section{Application to the Set Partition Problem}
\subsection{Statement of Problem}
Given a set $S = \{a_1,a_2,\ldots,a_n\}$ of positive numbers, find a subset
$s\subset S$ such that $P$ is minimized, where \begin{eqnarray} \label{eq:partition} 
P(s) = \Big| \sum_{a_j\in s} a_j - \sum_{a_k\notin s} a_k \Big| 
\end{eqnarray}
Not only is this problem at the heart of NP-completeness \cite{Garey97},
but it is framed in the manner of a binary optimization that minimizes a
cost function.  While the nested Grover search is the best result so far
for solving NP-complete problems, it relies on a structure that does not
exist in optimizations.  
\subsection{Solving Set Partition}
\subsubsection{Using the Algorithm} \label{sec:usealg}
The function $P$ in (\ref{eq:partition}) takes the place of the function $P$
used in (\ref{defP}).  The first register of our quantum computer will still be $n$
qubits.  If each of the numbers in the set $S$ has $b$ bits of precision,
then the second quantum register will have to be $\lceil b+\log{n}\rceil$
qubits in order to accomodate the largest possible value of $P(s)$, which
is $N=n2^b$.  If $b$ is large, this requirement can be significantly
relaxed.  The only strict requirement is that $\phaseop$ has to
distinguish between goods and nogoods.

The registers also are set-up such that each qubit $i$ in the 1st register
corresponds to a specific $a_i$ in $S$ ($\Rightarrow$ $P(s) = \sum (-1)^ia_i$).
The set partition problem has a degeneracy because $P(s)$ is the
absolute value of a difference, so in solving this problem with a quantum
algorithm I only consider a database of solutions where the smallest
number is in the subset $s$.  In principle, however, any $a_i$ could be
used for this purpose.  Also, this problem becomes deterministic at $n=4$,
so I only solve cases $n\geq 5$. 

Most importantly, I need to specify exactly how to reduce the size of the
database.  Suppose that the state $\ket{m}$ that we measure has $l$ 1's in
its binary representation.  This means that there are $n-l$ $a_is$ in the
subset $s$ and $l$ $a_is$ not in $s$.  The procedure is as follows: if
$l$ is even, then choose the smallest $a_i$ and call it $t$.  In $S$,
replace each $a_i$ not in $s$ with the difference $a_i-t$ and remove $t$
from $S$.
If $l$ is odd, then add $a_1$ to the group of $a_i's$ not in $s$ and use
the same procedure.  In order to solve a problem instance, $n-4$
of these decisions must be made, and with classical processing they can
tracked to give a solution of $n$ variables at the end of the iterations.

\subsubsection{Simulations}
My wish is not to submit these simulations as primary evidence that my
method works for solving NP-complete and optimization problems.  Rather,
the above sections contain enough information to see that this will be
true, and I will discuss some of these points in section \ref{sec:comments}.  However, the
complexity of this solution cannot be predicted analytically, so a
simulation helps to quantify a few examples of using this algorithm.

The simulations were set-up as follows: a random instance of the problem
was generated, and the probability of measuring each state was calculated
using (\ref{eq:probFN}).  Each possible measurement was tagged good or bad
based on whether it would led to inclusion of the $i_0$ after database
reduction.  For a given iteration on a specific problem instance, these
probabilities can be summed to obtain the total probability of making
either a good or bad measurement.  When these probabilities are averaged
over many problem instances they are denoted $p_G$ and $p_B$ respectively.  

Given average values of $p_G$ and $p_B$ for runs with 5 qubits up to $n$
qubits, the complexity in terms of number of oracle calls can be
calculated as follows:

The probability of finding the correct solution in a given run:
\begin{eqnarray} P_c = \Pi_{i=5}^n p_G(i) \end{eqnarray}
The algorithm will produce a solution after $n$ iterations.  However, if
we choose to only use certain types of states for measurement, multiple
oracle calls may be required for a given iteration.  
The average number of oracle calls for a run is:
\begin{eqnarray} \label{eq:nonexp} N_0 = \sum_{i=5}^n\frac{1}{p_G(i)+p_B(i)}
\end{eqnarray}
In order for the correct solution to be found after $r$ runs of the
algorithm, an incorrect solution must be found $r-1$ times in a row,
followed by a correct solution on the $rth$ try.  Therefore, the
complexity is given by:
\begin{eqnarray} Complexity & = & P_c\cdot N_0 + (1-P_c)\cdot
P_c\cdot(2N_0)+(1-P_c)^2\cdot P_c\cdot(3N_0) + \ldots\nonumber \\ 
& = & N_0P_c\sum_{k=1}^{\infty} k(1-P_c)^{k-1}\nonumber \\ Complexity & =
& \nonumber \end{eqnarray}
\begin{equation} \label{eq:complexity}\Big(
\sum_{i=5}^n\frac{1}{p_G(i)+p_B(i)}\Big) \Big(\Pi_{i=5}^n p_G(i) \Big) \Big(
\sum_{k=1}^{\infty} k(1-(\Pi_{i=5}^n p_G(i)))^{k-1} \Big)  \end{equation}
A few comments: \\
1) The form of $P_c$ implies that this complexity will grow exponentially
unless $p_G \to 1$ as $n\to\infty$.  \\ 
2) In calculating the infinite sum from simulation data, I just add terms
until they are below the threshold $10^{-4}$.  I checked this against smaller
thresholds and it does not appear to affect results. \\
3) The asymptotic value of the exponential part of the complexity can be
estimated by $O((\frac{1}{p_G(i)})^n)$ using the largest $i$ for which $p_G$
is known, but this will not
necessarily give the whole picture.  

\subsubsection{Results of Simulation}
At each iteration, an $n$ variable problem is reduced to an $n-1$ variable
problem, thus reducing the size of the database by a factor of 2.  If the
target solution remains in the database, then this was a successful
reduction.  Figure 1 shows the average probability of a
successful database reduction at various problem sizes.  The first
measurement scheme repeats the first three steps of section \ref{sec:alg}
until $\ket{m}=\ket{2^k+2^j}$.  The second scheme uses any
$\ket{m}\neq\ket{0}$ as a valid measurement.  As figure 1
clearly indicates, the first scheme is more successful.  Based on the form
of (\ref{eq:complexity}), improving $p_G$ can save an exponential number of
steps, so the first scheme is adopted in future simulations.  

In these first two simulations, the $\epsilon$ appearing in (\ref{eq:defS}) and
(\ref{eq:defTheta}) was not explicity used.  In order to test an ideal
case, I \emph{cheated} and flipped the phase of exactly half of the states
in the database.  However, it was shown in (\ref{eq:probFN}) that this is
not necessary.  In order to obtain \emph{realistic} complexity data, I henceforth
simulate the algorithm using a naiive method that flips all states whose
cost is below
$.29\sum_i a_i$.  In general, if we know what fraction of states we
want to flip, better methods than the naiive one I employ are available
using the density of states for the partition problem \cite{Smely}.

As stated in section \ref{sec:alg}, the algorithm requires $n$ iterations
to find a solution.  However, if the $l=2$ measurement scheme is employed,
some measurements will be thrown out, in which case extra oracle calls
must be made.  Figure 2 is a plot of data for (\ref{eq:nonexp}) with a
linear fit.  This shows that we have paid a small price by using the $l=2$
measurement scheme: instead of taking $n$ iterations to get a
solution we need $O(n)$.

The asymptotic behavior of the algorithm is related to the asymptotic
behavior of figure 1.  It is possible that as
$n\rightarrow\infty$, $p_G(n)\rightarrow .5$, which is equivalent to a
random decision (there is no reason to expect $p_G$ would go lower than .5).  In this
case, the asymptotic behavior would be no better than a classical search.
If the asymptotic behavior of $p_G\ge .5$, then of course we are in luck.
Of course, this cannot be determined from simulation.  But by examining
figure 1 I argue that if $p_G$ falls off slowly enough, then
we still benefit.  

Figure 3 is a comparison between three different
searches.  A random classical search will, on average, find a solution
after $2^{n-1}$ tries.  The data for this algorithm is based on
simulations using the naiive phase-flipping scheme described above and
plugged into \ref{eq:complexity}.  An unstructured Grover search requires
$2^{\frac{n-1}{2}}$ oracle calls (rotates towards 2 target solutions).  Comparisons to classical search and
further comments are given in section \ref{sec:comments}.

However, the data in figure 3 slightly misrepresents the new approach,
because although each iteration of the algorithm makes at least one oracle
call, these oracle calls use fewer than $n$ variables.  For example, 
assume for a moment that the oracle function takes $O(n^2)$ steps to
implement.  When the algorithm makes an oracle call on $n-3$ variables,
let us count that as only $(\frac{n-3}{n})^2$ of a call, because both the
classical and Grover unstructured searches always make oracle calls using
all $n$ variables.  Figure 4 shows data adjusted in this manner, and it
shows a noticeable improvement.

\section{Comments on this Algorithm} \label{sec:comments}
Since this is a new method, it may be necessary to explain the intuition
behind how it works.  

\subsection{Summary of Results}
The algorithm presented in section \ref{sec:alg} can be used in order to
perform structured database search.  It returns a solution
after $O(n)$ oracle calls, but (\ref{eq:complexity}) shows that the
probability of this being an optimal solution decreases exponentially.
Each oracle call is followed by a measurement, implying that quantum
coherence is necessary for only one oracle call at a time.
\subsection{A Quantum Heuristic}
\subsubsection{Description of the Heuristic}
We have a database of size $2^n$.  Each of the $2^n-1$ possible states
that could be measured in step 4 of the algorithm implies a subspace with
$2^{n-1}$ members (see section \ref{sec:choose}).  We are most likely to
explore a given
subspace if the number of good states included in that subspace
and number of good states not included in that subspace differ (see equation (\ref{eq:probFN})).  
\subsubsection{Meaning of Heuristic in $l=2$ example}
In solving the set partition problem, measurements of the form $\ket{2^k+2^j}$ have the simple interpretation 
that the database is reduced by placing two numbers from the set $S$ in
different groups.  If placing two numbers in different groups results in a
large number of either low-lying or non-low-lying solutions , then the
probability of following this path is high.  In fact, the data in figure 2 implies that in the set partition problem, states of
the form $\ket{2^k+2^j}\ket{0}$ are peaked in $\op{D}\ket{0}\ket{0}$
The fact that $p_G$ is greater than .5 (as can be seen
in figure 1) means that a subspace with a non-random
distrubtion of low-lying states has a better than $50\%$ chance of
containing target solution.
\subsubsection{Quantum Heuristic vs. Classical Heuristic}
Differencing heuristics already exist to solve partition problems
classically \cite{Korf98,Mertens99}.  The quantum heuristic is different
because it can take into account properties of a whole database for a
specific problem instance.
For example, putting the largest two numbers into different sets is a good classical
heuristic because it works well on \emph{most} problem instances.
However, the probability of putting the two largest numbers in different
sets in a given run of this algorithm is based on how many low-lying
solutions that action creates in the specific instance being solved. 
\subsection{This is not Amplitude Amplification} \label{sec:amp}
It is important to understand that this approach does not utilize
amplitude amplification as used in other algorithms \cite{Grover97,Brassard00,Hogg98}.  The states that are peaked in
$\op{D}\ket{0}\ket{0}$ represent subspaces of the database to be searched
on future iterations.  \emph{They no longer represent the states indexing
the database}. 
Furthermore, there is the following difference with amplitude
amplification: if a classical algorithm makes $2^n$ oracle calls, it will
have explicity checked the cost of $2^n$ different database items.
However, if a Grover search makes $2^{\frac{n}{2}}$ oracle calls to rotate
to a target state, at the end it will only have explicitly checked the
cost of a single database item because only one measurement occurs.  This
algorithm falls somewhere in between those two extremes.  After every $O(n)$
oracle calls a database item is checked.  However, this algorithm is more likely
than a classical algorithm to check the same database item multiple times
because quantum measurement is probabilistic.  
\subsection{Usefulness of this Approach}
The most important goal is to find a solution as quickly as possible.  So
far, I have not demonstrated that this algorithm is any better at finding
solutions than a Grover search (although it has different properties as
noted in section \ref{sec:amp}).  Whether or not there are benefits
reducing the required coherence time is hard to say.  If it becomes
more difficult to perform many logic gates as system size increases
(see \cite{Ladd00} for proposed implementation where this is true), then
it is possible to imagine a situation where it is much more feasible to
implement this algorithm than a Grover search for certain problem sizes.  

As compared to classical algorithms, this approach has interesting
properities.  It is certainly impossible to solve a problem this way
classically.  The steps in section \ref{sec:alg} outline what is probably
the simplest approach to using the properties of an entire database to
decide how to parse the tree of possible solutions to an optimization
problem.  It may be that this algorithm's usefulness would be not be in
solving the set partition problem, but in solving problems where little is known about
the database structure a riori.   
\section*{Acknowledgments}
This work has been supported by the Research Institute for Advanced
Computer Science at NASA Ames Research Center.  I would not have started
work on this without the help of Vadim Smelyanskiy and Dogan Timu\c{c}in.
I would also like to thank Cyrus Master for finding a crucial error
in the original version of this algorithm during a presentation at
Stanford last November.

\section*{Comments}
I would appreciate comments sent to brian34@feynman.stanford.edu.

\newpage

\begin{figure}
\label{fig:meas}
\begin{center}
\psfig{figure=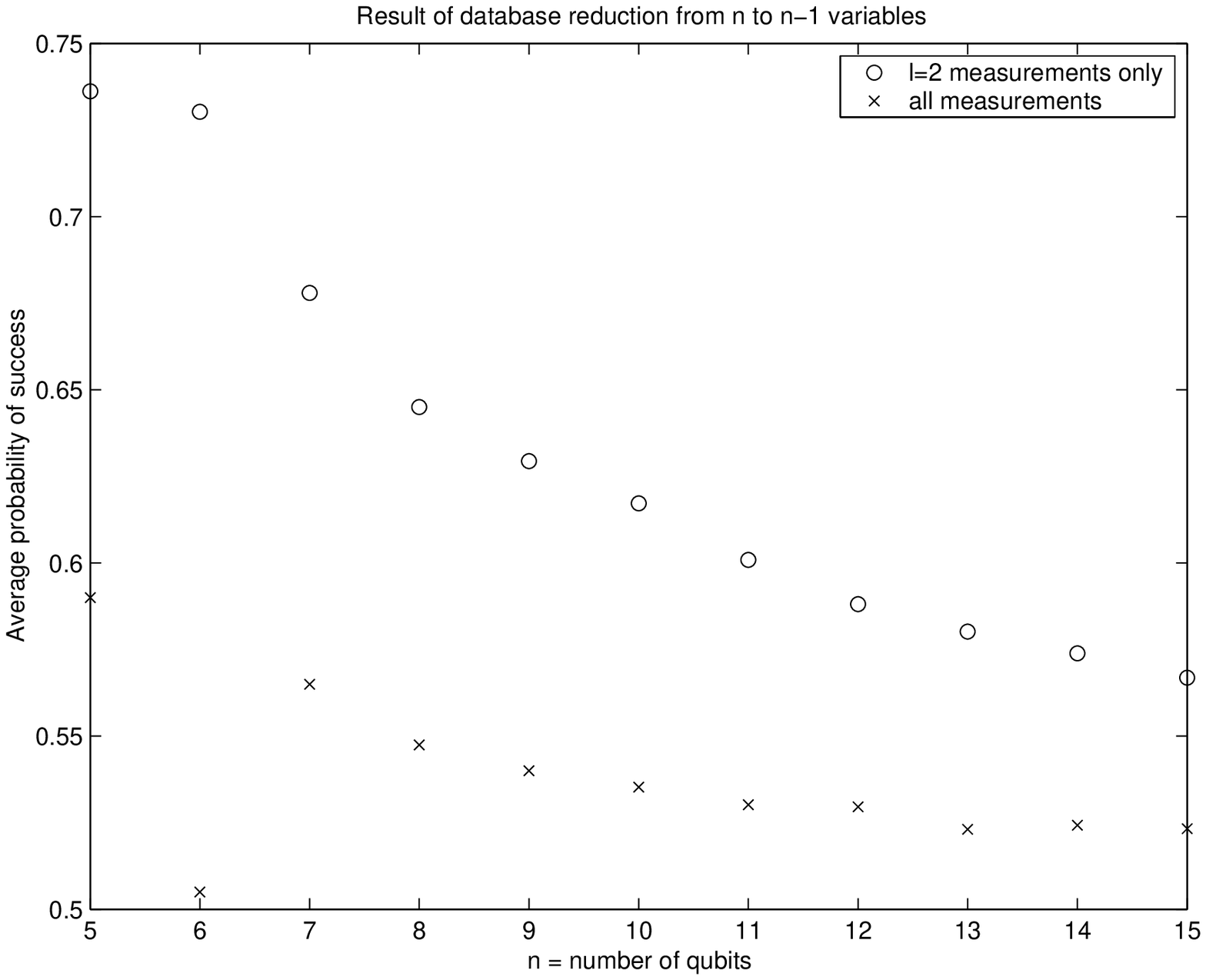,height=5in}
\end{center}
\begin{center}
\caption{The probability of a correct database reduction from $n$ to $n-1$
variables is shown for two schemes.  In one scheme $\ket{m}=\ket{2^k+2^j}$
is allowed.  In the second, $\ket{m}\neq\ket{0}$ is allowed.  In both
simulations, $\phaseop$ is idealized to flip the phase of half of the
states in the database.  Each data point in the first scheme is the
average of 100 problem instances, each data point in the second scheme is the average of 50 instances.}
\end{center}
\end{figure}

\begin{figure}
\begin{center}
\label{fig:nonexp}
\psfig{figure=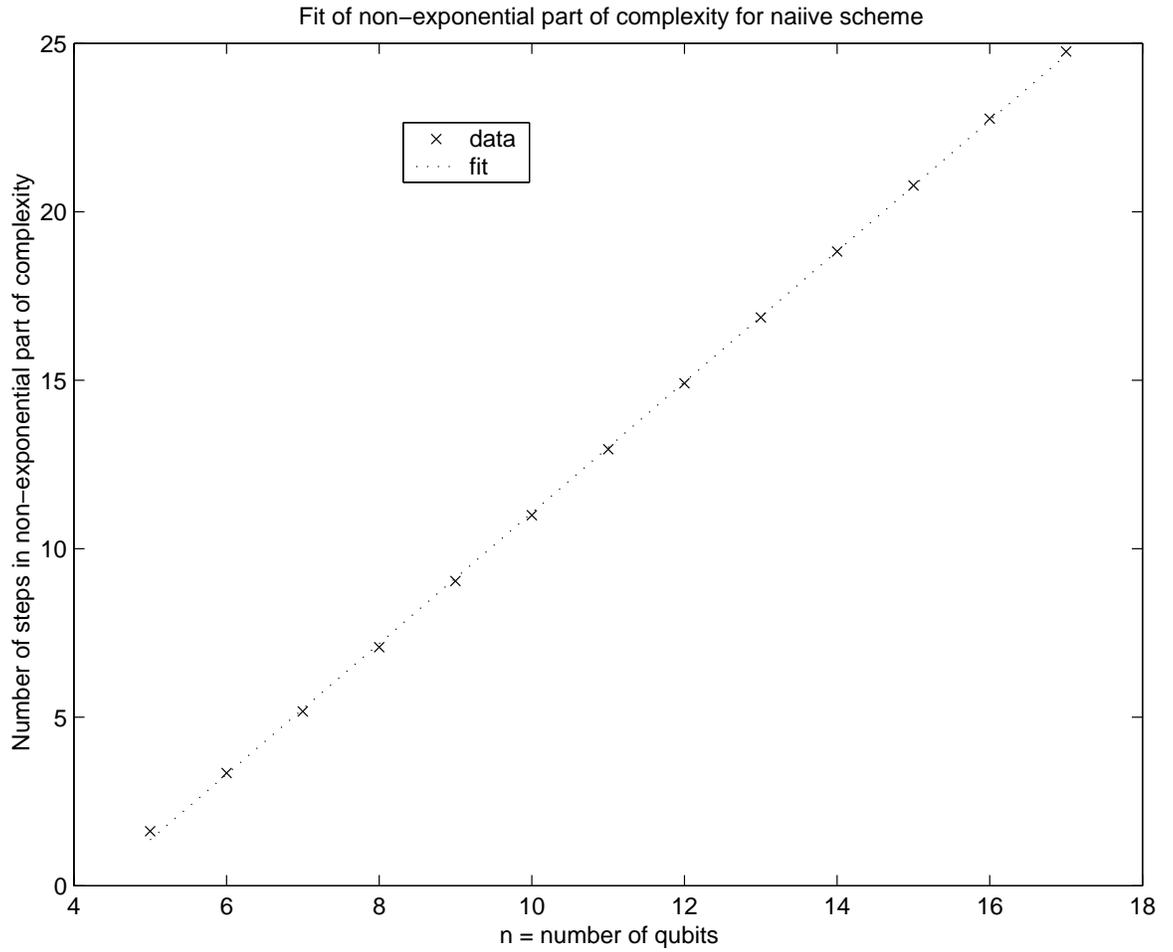,height=5in}
\end{center}
\begin{center}
\caption{Plot of the number of oracle calls necessary to perform one run
of the algorithm for a simulation with $\epsilon = .29\sum_i a_i$.
This data is plotted to quantify the 
extra runs that are required in order to obtain $l=2$ measurements.
The data is fit to
$1.9401\cdot (n-4) - .5735 $ to show that the number of calls per run is
still $O(n)$.} 
\end{center}
\end{figure}

\begin{figure}
\begin{center}
\label{fig:unstruc}
\psfig{figure=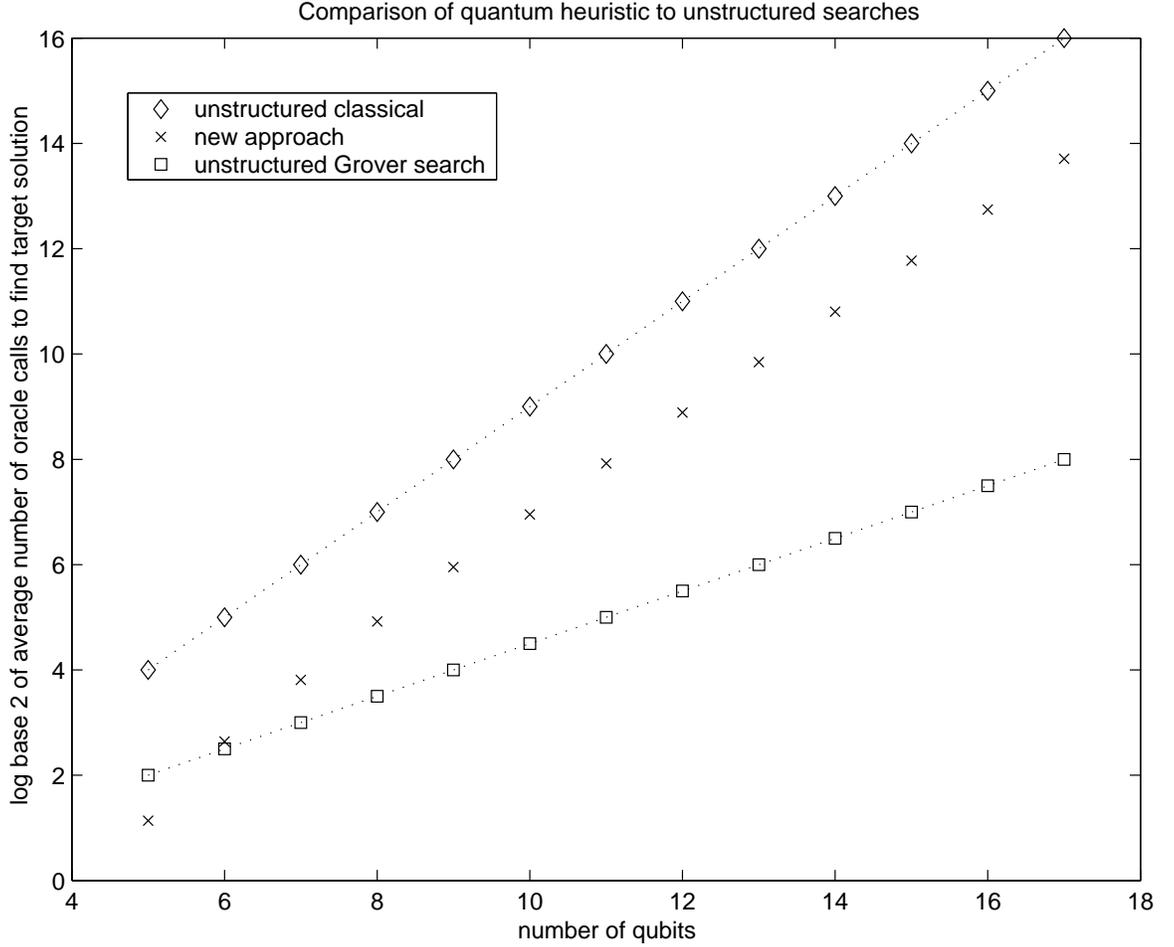,height=5in}
\end{center}
\begin{center}
\caption{Comparison between number of oracle calls for random classical
search ($2^{n-1}$), new algorithm (from simulation and equation
\ref{eq:complexity}), and unstructured Grover search ($2^{\frac{n-1}{2}}$).
Simulation data is average of 100 problem instances at each value of $n$ from 5
to 15 qubits, 50 instances at 16 qubits, and 40 instances at 17 qubits.
Instead of tracking one problem instance from beginning to end, average
probability of successful reduction from $n$ to $n-1$ variables is
empirically calculated and plugged into equation (\ref{eq:complexity}).} 
\end{center}
\end{figure}

\begin{figure}
\begin{center}
\label{fig:weight}
\psfig{figure=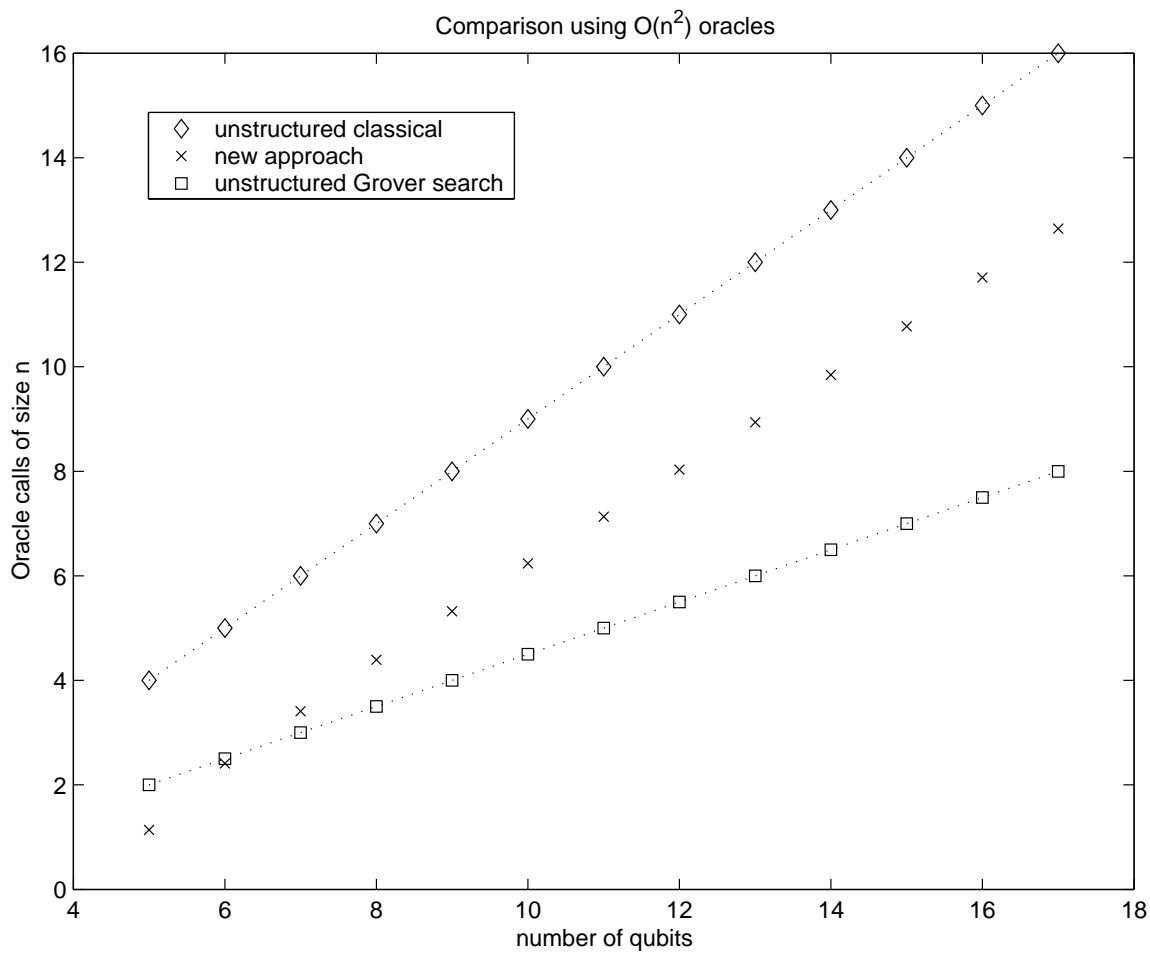,height=5in}
\end{center}
\begin{center}
\caption{Although the data presented in figure 3 is accurate in terms of
the actual number of oracle calls, this algorithm evaluates most of those
oracle calls on only a subset of the original variables.  Taking, for
example, an oracle that has complexity $O(n^2)$, we can adjust the data of
the new approach by weighing the oracle calls by how many variables they
operate on.  This shows a noticeable improvement.}
\end{center}
\end{figure}

\end{document}